# Universal Scaling Law of Glass Rheology

Shuangxi Song[1], Fan Zhu[1,3] and Mingwei Chen[2,3*]

[1]State Key Laboratory of Metal Matrix Composites, School of Materials Science and Engineering, Shanghai Jiao Tong University, Shanghai 200030, PR China

[2]Department of Materials Science and Engineering, Johns Hopkins University, Baltimore, MD 21218, USA

[3]WPI Advanced Institute for Materials Research, Tohoku University, Sendai 980-8577, Japan

*Correspondence to: mwchen@jhu.edu.

The similarity in atomic structure between liquids and glasses has stimulated a long-standing hypothesis that the nature of glasses may be more fluid-like, rather than an apparent solid [1-5]. In principle, the nature of glasses can be characterized by measuring the dynamic response of rheology to shear strain rate in the glass state. However, limited by the brittleness of glasses and current experimental techniques, the dynamic behaviours of glasses were mainly assessed in the supercooled liquid state or in the glass state within a narrow rate range. Therefore, the nature of glasses has not been well elucidated experimentally. Here we report the dynamic response of shear stress to shear strain rate of metallic glasses over nine orders of magnitude in time scale, equivalent to hundreds of years, by broadband stress relaxation experiments. The full spectrum dynamic **response of metallic glasses, together with other "glasses" including silicate and polymer glasses, granular materials, soils, emulsifiers and even fire ant aggregations, follows a universal scaling law within the framework of fluid dynamics. Moreover, the universal scaling law provides comprehensive validation of the conjecture on the jamming phase diagram by which the dynamic behaviours of a wide variety of "glass" systems can be unified under one rubric parameterized by thermodynamic variables of temperature, volume and stress in the trajectory space.**

It has been well established that plastic flow of crystalline materials is predominantly driven by the formation/motion of discrete crystal defects such as vacancies, dislocations, stacking faults, twins and grain boundaries in various stress, time and temperature domains [6]. In contrast, simple and complex liquids behave in a viscous or viscoelastic fashion. While, unlike crystals and liquids, glasses, such as metallic glasses (MGs) and window (oxide) glasses, usually have liquid-like disordered atomic/molecular structures but are stiff and hard as brittle solids. It has been long conjectured that glasses may retain the nature of liquids as suggested by some peculiar glass flowing observations from thickened bottoms of stained glass panes in old European churches [7]. Physically, the nature of glasses can be characterized by correlating the dynamic response of plastic flow to applied force in a wide time range. However, in the accessible timescale of conventional mechanical testing techniques, glasses usually behave as the elastic-plastic materials with minor temperature and strain-rate sensitivity. As a result, the comprehensive assessments of the dynamic behaviour of glasses have not been realized by direct experimental measurements or Williams–Landel–Ferry (WLF) time-temperature superposition function analysis in the glass state [8]. Under quasi-static loading conditions, plastic rheology of glasses is usually associated with the collection of local shearing events, most likely from small scale structure and density fluctuations [9,10]. For MGs, the local shearing is described by a dynamic variable, shear transformation zone (STZ), which represents the evolution of shear-induced local configurations [11-16]. The experimentally accessible parameter, which can microscopically depict plastic flow of MGs, is the statistical size of STZs. Depending on theoretical models and experimental methods, the estimated STZ volumes have been reported to vary over two orders of magnitude from several atoms to hundreds of atoms [17-20], which is far beyond possible experimental errors. In fact, the large discrepancy in the STZ volumes has caused intense debate and confusion on the understanding of rheology mechanisms in MGs. We noticed that previous experiments were carried out within a limited strain rate range in different time domains. The dispute of rheology behaviour may arise from a distinct rate/time dependence of flow mechanisms, although mechanical behaviour of MGs is usually considered time insensitive at temperatures far below glass transition points ($T_g$). Thus, to have a thorough understanding on the rheology and thus the nature of the hard glasses, it is necessary to design and carry out experiments that are capable of measuring the dynamic response of MGs over a broad strain-rate range.

As shown in **Figure 1a** we measured the dynamic shear stress response of a $Zr_{55}Cu_{30}Al_{10}Ni_5$ MG [21] at shear strain rates ranging from $10^{-8}$ to $10^0$ s$^{-1}$ by combining broadband nanoindentation and cantilever bending methods at room temperature. In the stress relaxation experiments, shear strain rate ($\dot{\gamma}$) and nominal shear stress ($\tau$) are derived from the contact force ($P$) and displacement ($h$) at different relaxation time ($t$) (see **Methods**). The estimated shear stress is plotted as the function of shear strain rate for the high-rate nanoindentation (**Fig. 1b**) and low-rate cantilever experiments (**Fig. 1c**). The stress relaxation experiments at low-stress levels are essentially the creep testing. The small sampling volumes required by nanoindentation and cantilever experiments effectively avoid the interference of discrete shear banding and cracking at high-stress levels. The experimental data points obtained by two different methods perfectly overlap with each other in the strain rate range between ~$10^{-6}$ and $10^{-5}$ s$^{-1}$ and form one completely smooth curve (**Fig. 2a**). For the first time, the dynamic mechanical response of a material in the glass state is presented in such a wide rate range from ~$10^{-8}$ to $10^{0}$ s$^{-1}$, spanning nine orders of magnitude in time scale. In comparison with the conventional dynamic mechanical analysis (DMA) by which a wide time range of elastic and viscoelastic response can be obtained by the time-temperature superposition [22], the broadband stress relaxation experiment allows us to directly measure the stress-strain responses at a temperature far below T$_g$ and a stress above yield point. The capability of low temperature and high stress measurements is essential for illustrating the full-spectrum dynamic response of the glass state as shown in **Fig. 2a**. By converting the stress-strain relation into a log-log plot, there are two distinct linear regions with two dramatically different slopes, *i.e.* strain rate sensitivities ($m = \partial \ln \tau / \partial \ln \dot{\gamma}$), at high and low strain rates (stresses), respectively. At high strain rates ($> 10^{-4}$ s$^{-1}$), the experimental data can be well fitted by Johnson-Samwer cooperative shear model (CSM) portrayed by a sinusoidal function of megabasin potential energy density [13]:

$$\dot{\gamma} = \dot{\gamma}_0 \exp\left(-\frac{W_{CSM}}{kT}\right) \quad (1)$$

where $\dot{\gamma}_0 \approx 10^{10}$ s$^{-1}$ is the elastic strain rate limit [13], $W_{CSM}$ is the activation energy of cooperative shear, which depends on stress (*see* **Methods**); $k$ is the Boltzmann constant and $T$ is temperature. On the other hand, at the strain rates below $10^{-6}$ s$^{-1}$, the flow rate follows a power law with a rate sensitivity or power index of unity. In this case, the rate-stress dependence can be described by the reversible STZ-free volume model originally derived from the transition state theory (TST) by Spaepen and Argon [11,12]:

$$\dot{\gamma} = \alpha \gamma_0 \nu_G \exp\left(-\frac{W_{STZ}}{kT}\right) \sinh\left(\frac{\tau \gamma_0 \Omega}{2kT}\right) \quad (2)$$

where $\alpha$ (of order unity) is the steady-state volume fraction of flow units contributing to plastic flow; $\gamma_0$ represents the transformation strain of STZs and is taken as 0.125; $\nu_G \approx 10^{13}$ s$^{-1}$ is the intrinsic phonon relaxation time [23]; $W_{STZ}$ and $\Omega$ are the activation energy and volume of STZs [11]. The activation energy and size of STZs by fitting the strain rate data in **Fig. 1d** and **2a** using **Equations (1)** and **(2)** are listed in **Supplementary Table 1**. The STZ size at high stresses is determined to be ~5.64 nm$^3$ or 338 atoms by CSM, consistent with previously reported rate-jump measurements of Zr-based metallic glasses at high stresses and strain rates [18]. The activation energy for the cooperative shear is estimated to be 5.91 eV, close to that for the α relaxation [24]. In contrast, the STZ volume at low stresses is measured to be only ~0.22 nm$^3$ or 16 atoms at low stresses, in agreement with previous relaxation spectrum measurements [17] as well as those measurements at temperatures close to $T_g$ [20]. Correspondingly, a low activation energy of only ~1.08 eV and small activation volume ($V^*$) of 0.028 nm$^3$ or a single atom by $V^* = \gamma_0 \Omega$ are obtained from the TST equation. Both the activation volume and activation energy indicate that dynamic flow of the MG at low strain rates can be described by single-atom motion or synchronized motion of all atoms at the same pace. In particular, the linear correlation between shear stress and shear strain rate at rates below $10^{-7}$ s$^{-1}$ indicates that the MG follows the behaviour of a Newtonian flow at low rates.

In the transition region, the strain rate sensitivity drops sharply from ~1.0 to 0.004 within a narrow strain rate range between $10^{-6}$ and $10^{-4}$ s$^{-1}$ (inset in **Fig. 2a**). It suggests a dynamic transition from viscoelastic to elastic-plastic deformation with the significant increases in flow stress, activation energy and STZ volume. We noticed that the dynamic transition includes two distinct stages: from Newtonian to non-Newtonian and from the Non-Newtonian to cooperative shear, as marked by the arrows in **Supplementary Figure 1a**. Although Johnson-Samwer CSM model [25] and Langer-Falk STZ model [12] predict the Newtonian to non-Newtonian and/or the Non-Newtonian to cooperative shear transitions, there is obvious divergence between the broad-band experimental data and these models (**Supplementary Figure 1a and Supplementary Figure 7**). In fact, we found that the experimental data in the transition region can be best fitted by Kohlrausch (or stretched exponential) relaxation function as shown by the black line in **Fig. 2a** and **Supplementary Figure 1.** This indicates that the dynamic response in the transition region may involve many-body interactions [26,27], most likely related to the thermal and athermal coalescence of STZs, flow units [22] or shear-microdomains [28]. It is in great contrast to the CSM and STZ models [8-10, 12] as they essentially describe a single body process.

On the basis of the shear stress and shear strain rate measurements in **Fig. 2a**, the viscosity ($\eta = \tau/\dot{\gamma}$) of the MG is calculated and plotted as the function of strain rate in **Supplementary Fig. 2**. The Newtonian viscosity ($\eta_N$) of the MG is estimated to be ~$8.5 \times 10^{14}$ Pa·s at room temperature. The viscosity of the MG at room temperature and literature data of other Zr-based MGs achieved at high temperatures near $T_g$ [29] are plotted in **Supplementary Fig. 3a**. The viscosity data are normalized by the Newtonian viscosity at each temperature and plotted as a function of strain rate in **Supplementary Fig. 3b**. The relationship between normalized viscosity ($\eta/\eta_N$) and strain rate of all data show the same trend from Newtonian flow at low strain rates to shear thinning at high strain rates. These observations are in good agreement with previous viscosity measurements and simulations of glass-forming liquids [25,29-31]. The normalized viscosities of MGs are plotted as a function of the scaled shear strain rate ($\dot{\gamma}\eta_N$), *i.e.* viscous stress in Newtonian flow (**Fig. 2b**), which includes the data from the current broadband experiments and the literature data from high-temperature deformation, room-temperature creep and DMA of Zr-based and La-based MGs [29,32-35]. It is immediately evident that all the data approximately fall on one master curve that spans a wide shear strain rate regime from $10^{-9}$ to $10^6$ s$^{-1}$, independent of temperatures (below or above $T_g$, i.e. glass or liquid states), chemical composition and sample geometry.

We noticed that the similar dynamic trend of normalized viscosity as the function of scaled strain rate can be obtained from a wide range of "glass" systems, including inorganic glasses [36], polymer glasses [37-39], emulsifiers [40], granular materials [41], soil liquefaction [42], and even fire ant aggregations [43] (**Fig. 2c**). However, the viscous stresses (i.e. $\dot{\gamma}\eta_N$) of these "glass" systems vary several orders of magnitude. Based on the scaling law for the yield strength of glasses, $\tau_C = 3k(T_g - T)/V$, acquired in our previous study [44], the normalized viscosity is further plotted as the function of a dimensionless parameter $\dot{\gamma}\eta_N \dfrac{V}{3kT_g}$ in **Fig. 2d**, where $V$ is the average molar volume determined from average atomic or particulate weight $M$ and density $\rho$, i.e. $V = M/\rho$ (see **Supplementary Table 2**). It is remarkable that the experimental data from all "glass" systems fall within a close proximity of one master curve in **Fig. 2d**, demonstrating that the dynamic response of the "glass" systems can be described by a universal scaling law. In fact, this is the first time that the rheology of *a wide variety of disordered systems is unified by a single scaling law*. Note that $\dot{\gamma}\eta_N \dfrac{V}{3kT_g}$ is equivalent to $\dfrac{\dot{\gamma}\eta_N}{\tau_C} = B_i$, where $B_i$ is the Bingham number defined as the ratio of the viscous stress ($\dot{\gamma}\eta_N$) of liquids to the elastic shear stress limit (*i.e.* yield strength) $\tau_C$ at $T = 0$ K with a molar volume $V$ [45]. Therefore, *the universal scaling law that is derived to describe the* *dynamic response of the "glass" systems is, in fact, within the framework of the classical fluid dynamics.*

The scaled rheological behaviour exhibits two linear regimes with distinct slopes in the double-log plot (**Fig. 2d**). The slope is zero in the low-rate Newtonian flow region, suggesting a rate-independent property of equilibrium liquids. By contrast, in the high flow rate or high viscous stress region, the slope gradually approaches to -1. Similar to the ideal steady flow of polymeric liquids at high strain rates [46], the negative unit slope in the non-Newtonian regime indicates that the nominal shear stresses of the elastic-plastic behaviour reach the maximum and the plastic flow is dominated by an athermal process. It is evident that, if the applied stress or experimental time (strain rate) is below a critical value, *i.e.* $\dot{\gamma}\eta_N \frac{V}{3kT_g} < 1$, the "glass" systems show the liquid behaviour and the rheology is dominated by thermally activated Newtonian flow (**Fig. 2d**). However, when $\dot{\gamma}\eta_N \frac{V}{3kT_g} > 1$, the systems become solid-like and the rheology is mainly controlled by stress-driven cooperative shear which is manifested as the coalescence of STZs, flow units [22] or shear-microdomains [28]. Conventionally, this yielding or shear thinning process is considered as the onset of time-independent elastic-plastic flow in glassy materials or unjamming in granular materials as the transient plastic flow is readily noticed. In other words, the universal scaling law reveals a mechanistic transition from thermally activated Newtonian flow to stress-driven cooperative shear which takes place when

$$\dot{\gamma}\eta_N \frac{V}{3kT_g} = 1 \qquad (3)$$

Therefore, the transition between liquid and glass essentially depends on four parameters: molar volume $V$ or density; temperature, $T$; experimental or observation time, $\dot{\gamma}$ or $t_{obs}$; and stress, $\tau$. In principle, the transition can be defined by one of the four variables and described by a three dimensional (3D) contoured surface in temperature-stress-volume, temperature-stress-time, temperature-time-volume, or time-volume-stress phase spaces. For example, in a typical case of time being the critical variable [32,47-49], the transition can be presented on a 3D contoured surface in the temperature-stress-volume space with three scaled variables of $kT/U_0$, $\tau/\tau_0$ and $V/V_0$ as the axes (**Fig. 3a**). Here, $U_0=3kT_g$ is the thermal energy required for glass transition at zero stress and the maximum density [44], $\tau_0$ is the critical strength at 0 K and the maximum density, and $V_0$ is the molar volume at 0 K and at zero stress. The transition boundary can be determined by critical-like equations where the three scaled variables are correlated [48,50] (*see* **Supplementary Information**). Experimental data of the transition for metallic glasses, plotted as the red solid spheres, locate at the transition boundary [13,51], together with the literature data from a granular system under a constant driving force or tapping as the dark grey spheres [52] and the attractive particles with $U_0 \approx 20kT$ as the navy spheres [48] in **Fig. 3a**. The transition points from the master curves in **Fig. 2d** are also plotted in **Fig. 3a**. The glass systems are in an equilibrium and reversible state underneath the contoured boundary, but become off-equilibrium and irreversible above the boundary.

In practice, one of the three thermodynamic variables often keeps constant or approximately unchanged during experiments and thereby the transition can be determined by a planar cut of the 3D contour. For examples, under a constant driving force, the transition boundary in the granular system from reversible to irreversible regimes by jamming and unjamming can be plotted as a contour curve in the 2D domain of configurational temperature (or tapping amplitude) and molar volume (or density) in **Fig. 3b**. For the attractive particles at a constant temperature, the transition boundary is a contour curve in the $V/V_0$ (or $(V/V_0)^{\lambda_V}$) and $\tau/\tau_0$ domain in **Fig. 3c**. The exponential dependence is related to the formation of attractive particle clusters and the density of a fractal aggregate decreases as the cluster grows [48,53]. In the case of metallic glasses, the density change is negligible and the transition is well presented in the temperature-stress domain in **Fig. 3d**. When the stress is small or remains nearly constant in conventional viscosity measurements, the dynamic transition is presented as the correlation between time and temperature, similar to the observed crossover from Arrhenius to non-Arrhenius liquids [54,55]. For our broadband stress relaxation measurements, the temperature and volume are nearly constant. Thus, the rheological behaviour is degenerated to the correlation between time (strain rate) and shear stress. As a result, the transition can be conveniently described by either correlated time (strain rate) or stress [32,44].

Liu and Nagel [47] have proposed a famous conjecture that the dynamic transitions in a variety of disordered systems, such as jamming-unjamming transition in granular materials and foams and liquid-to-glass transition in glass-forming materials, can be brought together under one rubric and unified by a 3D jamming phase diagram parameterized by thermodynamic variables of temperature, density (volume) and force. In the past decade, this hypothesis has been partially confirmed by computer simulations and experiments in granular and glass systems [32,48,49,56,57]. However, the comprehensive validation has not been achieved. It is intriguing to note that our universal scaling law, derived from the dynamic responses of a broad range of "glass" systems, provides theoretical confirmation on the hypothesis. The resulting phase diagram in the temperature-stress-volume space (**Fig. 3**), identical to the concept of jamming phase diagram, can quantitatively represents the experimental observations on temperature- and volume-controlled jamming transition of a granular system with constant driving force, stress- and volume-controlled jamming transition in attractive particles, and temperature- and stress-controlled glass transition in glass materials. Importantly, the unified phase diagram from the universal scaling law unveils the dynamic nature of "glass" systems in the framework of liquid dynamics, *i.e.* the jamming and glass transitions occur in trajectory space (space-time) rather than in configuration space and, therefore, are controlled by thermodynamic variables of stress (pressure), volume (density) and temperature that drive the systems out of equilibrium [58]. In fact, the "glass" systems are ergodic when they are in the equilibrium Newtonian flow region. When the system is out of equilibrium above the phase boundary, it is nonergodic and the motions of the constituent particles become correlated throughout observation time. The boundary for the transition where the ergodic and nonergodic phases are separated by a critical temporal variable can thus be described by a contoured surface of three thermodynamic parameters of stress (pressure), volume (density) and temperature by the universal scaling law and the dynamic (jamming) phase diagram.

Besides the scientific importance of the universal scaling law in unveiling the nature of "glasses" and in describing the dynamic transitions of "glass" systems, the scaling law also provides important guidance for designing and employing "glass" materials for practical applications. For example, the equal role of temperature and stress in glass transition directs the alloy design of high-temperature metallic glasses by considering strong and refractory elements [59], and warrants attention on the thermal and mechanical stability of glasses as a force-bear component serving at the temperatures even below $T_g$ that is conventionally determined by differential scanning calorimetry at zero force. For granular materials, the universal scaling law may bring new insights and attention on the possible effect of configurational temperature, in addition to the known pressure and density, on landslides and riverbank collapses by soil liquefaction[60].

In summary, the broadband dynamic response of MGs, together with a wide variety of other "glass" systems, can be unified by a universal scaling law within the framework of fluid dynamics. The scaling law provides comprehensive validation of the conjecture of jamming phase diagram and demonstrates that the dynamic transitions of disorder systems can be described by the conversion between equilibrium Newtonian liquids and off-equilibrium elastic-plastic solids. This work uncovers the liquid nature of glasses and offers quantitative description on the dynamic transitions of "glass" systems by thermodynamic variables of temperature, volume and stress in trajectory space.

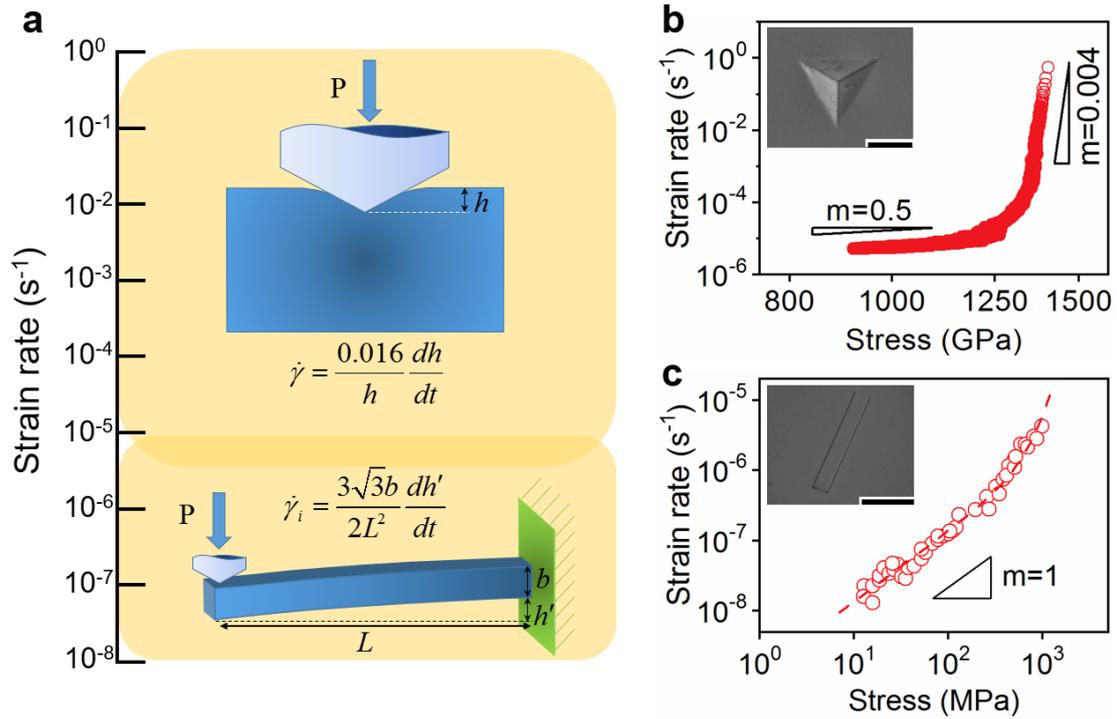

**Fig. 1. Broadband relaxation experiments of a Zr$_{55}$Cu$_{30}$Ni$_5$Al$_{10}$ Glass. a**, Schematic drawing of broadband stress relaxation experiments by using nanoindentation (above) and cantilever bending (bottom) tests with a total measurable strain rate range from 10$^{-8}$ to 10$^0$ s$^{-1}$. **b, c,** Shear strain rate as a function of shear stress calculated from nanoindentation relaxation measurements (**b**) and from cantilever bending relaxation measurements (**c**). SEM images of an indent after a nanoindentation relaxation test at 10 mN for 2000 s with a scale bar of 1 μm and a typical cantilever with a scale bar of 100 μm are inserted respectively. Rate sensitivity (*m*) of 1, 0.5 and 0.004 is plotted as triangle slope for eye guide.

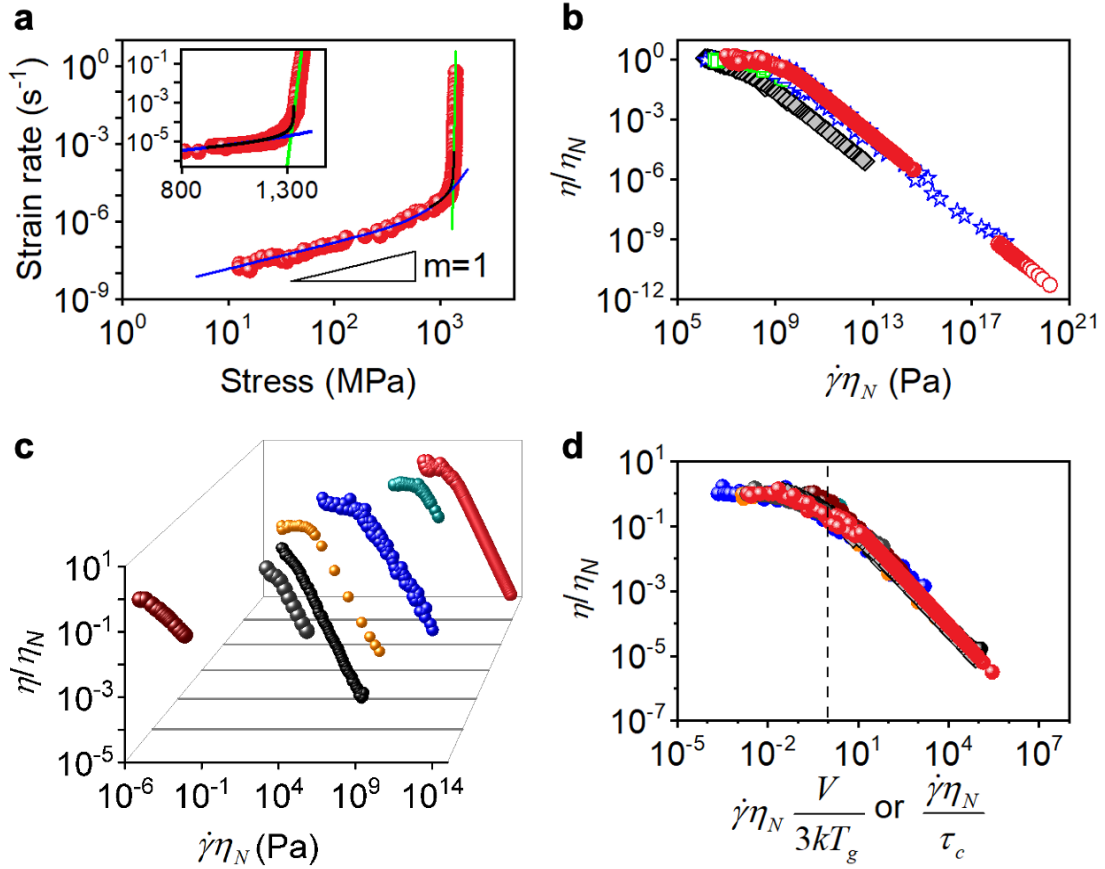

**Fig. 2. Strain rate-stress behaviour and universal scaling law of glass rheology. a,** The plot of logarithmic shear strain rate as a function of logarithmic shear stress for the $Zr_{55}Cu_{30}Ni_5Al_{10}$ glass under nanoindentation and cantilever bending at room temperature. Rate sensitivity (*m*) of 1 is plotted as triangle slope for eye guide. The inset is the enlarged view of the transition segment together with the fitting curves based on the transition state theory (blue line), cooperative shear model (green line), and Kohlrausch relaxation equation (black line). **b,** Normalized viscosity as a function of scaled strain rate by Newtonian viscosity from the broadband relaxation measurements at room temperature (red balls), DMA relaxation measurements on $La_{60}Ni_{15}Al_{25}$ metallic glass at various temperatures with 0.005 initial strain (grey diamonds) [22], creep data from compression at room temperature

(blue open triangles) [33,35] and 613 – 683 K in Vitreloy 1 glass (green open squares) [29] together with molecular dynamic simulation results in a Cu-Zr glass (blue pentacle) [32] and viscosity measured for propagating shear band in a Zr-BMG at room temperature (open red circles) [34]. **c**, Normalized viscosity as a function of scaled strain-rate by Newtonian viscosity from various glassy materials. Metallic glasses data is plotted as red balls. Silicate glass non-Newtonian flow data is plotted as dark cyan balls [36]. Viscosity data in polyethylene[38], polypropylene[37], and polybutadiene rubber [39] are plotted as blue balls. The ant aggregation viscous data is plotted as dark balls [43]. Mayonnaise non-newtonian flow data is plotted as orange balls [40]. Granular [41] and soil [42] data are plotted as dark grey and wine balls. **d**, Normalized viscosity as a function of dimensionless quantities as $\dot{\gamma}\eta_N \frac{V}{3kT_g}$ or $\frac{\gamma\eta_N}{\tau_c}$ in various glassy materials and the dashed line at $\dot{\gamma}\eta_N \frac{V}{3kT_g}=1$ is for guidance.

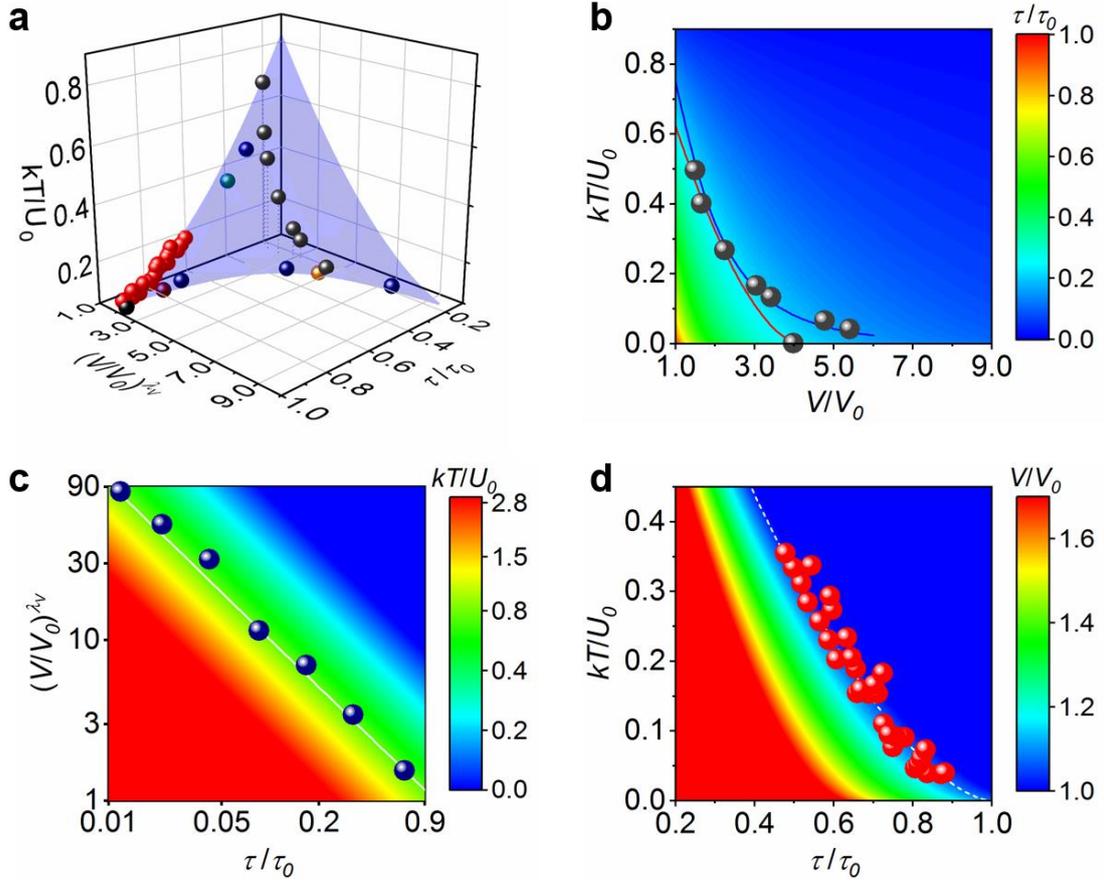

**Fig. 3: Dynamic transition phase diagram based on the universal law derived from the broadband relaxation measurements. a**, Perspective view of the transition boundary diagram plotted as light blue surface with normalized stress $\tau/\tau_0$, normalized molar volume $(V/V_0)^{\lambda_V}$, and reduced temperature $kT/U_0$ following **Equations (3)** and **(S15)** when $\lambda_V = 1$. The temperature dependence on $\tau/\tau_0$ and $V/V_0$ in various metallic glasses [13] and this study is plotted as red balls which coincide on the boundary surface at the MG region. The jamming transition boundary in attractive particles from literature at $U_0 \approx 20kT$ [48] and granular materials [52] are plotted as navy balls and dark grey balls respectively. The transition points of the glassy materials from **Fig. 2c** are also plotted. **b**, Phase diagram is plotted of two control parameters: $kT/U_0$ and $V/V_0$ in a granular system with constant driving force. **c**,

Two-dimensional boundary log-log plot of $V/V_0$ as a function of $\tau/\tau_0$ at $U_0 \approx 20kT$ for attractive polystyrene particles, the molar volume is normalized with $(V/V_0)^{\lambda_V}$ with $\lambda_V = 3.2$. **d**, Two-dimensional boundary plot of $kT/U_0$ as a function of $\tau/\tau_0$ at constant $V/V_0$ for metallic glasses. See **Supplementary Information** for details on curve fitting.

**Methods**

**Sample preparation**

$Zr_{55}Cu_{30}Al_{10}Ni_5$ (at %) metallic glass is selected as a model system for the broadband stress relaxation experiments. Glassy films with 5 μm thickness were prepared by a single-target magnetron sputtering method using a multicomponent $Zr_{55}Cu_{30}Al_{10}Ni_5$ alloy as the target [21]. The deposition was conducted with a radio frequency power of 100W at room temperature. The base pressure was $3\times10^{-4}$ Pa and the working pressure was tuned to 0.3 Pa with high purity argon gas. Si (100) wafers with diameter of 50 mm were used as substrates. Freestanding film samples were obtained by dissolving the silicon substrate in NaOH-water solution. The glass transition temperature of the film was measured to be 695 K by DSC with a PerkinElmer 8500 instrument under a purified argon atmosphere at a heating rate of 20 K/min, consistent with that of the $Zr_{55}Cu_{30}Al_{10}Ni_5$ bulk samples and previous reports [21,61]. The 5 μm thick, 25 μm wide, and 200 μm long cantilevers were machined by focused ion beam (FIB) in a dual beam SEM (FEI Versa 3D) system with 30 kV accelerating voltage and 0.5 nA $Ga^+$ ion current. Larger cantilevers of 30 μm thick, 70 μm wide and 700 μm long are prepared by machining $Zr_{55}Cu_{30}Ni_5Al_{10}$ ribbons using a low speed saw. Ribbon samples were produced by melt-spinning the $Zr_{55}Cu_{30}Al_{10}Ni_5$ alloy on a copper wheel at a rotation speed of about 420 rad·$s^{-1}$ in argon atmosphere.

**Structural characterization**

The microstructure of the as-prepared thin films was characterized by transmission electron microscopy (TEM). The TEM samples were prepared by ion milling using a Gatan PIPS-II ion miller equipped with liquid nitrogen cooling holder. TEM experiments were conducted by a JEOL ARM200F TEM operating at an acceleration voltage of 200 kV. The selected area electron diffraction pattern of the $Zr_{55}Cu_{30}Al_{10}Ni_5$ thin films shows diffusive amorphous halos in **Supplementary Fig. 4a**, demonstrating the amorphous nature of the samples. The high-resolution TEM image of **Supplementary Fig. 4b** further verified that visible crystallites could not be found and the alloy used in the stress relaxation experiments has a fully amorphous structure. The morphology and size of the cantilever samples and the indents of nanoindentation experiments were inspected by scanning electron microscope (SEM) in a FEI Versa 3D SEM with 5 kV accelerating voltage.

**Broadband nanoindeantion relaxation experiments**

The elastic modulus (E) and hardness (H) of the metallic glass is determined to be 121.5 GPa and 6.7 GPa at room temperature (295 K), following Oliver-Pharr method [62], by a nano-indentation instrument (Hysitron Triboindenter, Minneapolis, MN) equipped with a Berkovich indenter at the peak force of 10 mN and the loading/unloading rate of 0.2 mN/s. The nanoindentation relaxation tests were performed by applying the load to 10 mN in 0.05 seconds, holding at 10 mN for up to 200 seconds for quasi-static relaxation measurements and up to 2000 seconds for dynamic relaxation measurements, and then unloading in 0.05 seconds. Indentation depths were all within 300 nm, much smaller than 10% of the film thickness, avoiding possible influence from substrates. The load and depth were recorded at data acquisition rate up to 2000 points per second for relaxation tests.

In order to obtain reliable and accurate relaxation data, the following procedures are followed. The area function of the probe is carefully calibrated on fused quartz to get area function constants $C_1$ to $C_6$ as listed in **Supplementary Table 3** by **equation (S1)** [62].

$$A(h) = \sum_{i=1}^{6} C_i \cdot (h - 0.75 P_{\max}/S)^{4/2^i} \quad \quad (S1)$$

where $A(h)$ is the contact area as a function of indentation depth ($h$), $P_{\max}$ is the peak force, and $S$ is the contact stiffness measured from unloading segments. The nanoindentation relaxation results in this study were obtained at depths larger than 200 nm to avoid the tip radius effect which is evident at depths lower than 30 nm. The geometry of the indents for 2000 seconds relaxation tests is also characterized by SEM and is shown in **Fig. 1b**.

In order to minimize the thermal drift effect from the nanoindentation relaxation measurements, all tests were performed at least 2 hours after samples were placed in the instrument chamber for sufficient system stabilization. The typical thermal drift rate of the system was calibrated by a fused quartz sample with 1 μN contact force in every 40 seconds and was plotted in **Supplementary Fig. 5a**. The thermal drift rates are all within ±0.02 nm·s$^{-1}$ in this study. Before each test, the probe is engaged onto the sample surface and waits 300 seconds for stage motor and piezo to settle. After that, the thermal drift was monitored for 100 seconds with 1 μN contact force before each test and with 5% peak force after each test. The average thermal drift rate was calculated and subtracted from each relaxation measurement. Typical quasi-static nanoindentation relaxation data with constant 10 mN force is plotted in **Supplementary Fig. 5b**. The measured strain rates are all above the thermal drift rate during 200 s relaxation. Thus, standard quasi-static nanoindentation relaxation measurement is reliable as long as 200 seconds, beyond which the thermal drift may interfere the low rate measurements. Additional indentation relaxation at lower stress and lower strain-rate was measured by loading to 10 mN and then monitoring the dynamic contact stiffness with 0.2 mN load amplitude and 200 Hz frequency for as long as 2000 seconds. The indentation depth was estimated from the measured dynamic contact stiffness which had negligible thermal drift effect because the high dynamic loading frequency is comparable to the low drift rate. The measured contact stiffness ($S$) was used to estimate the indentation depths ($h$) following **Equation (S2)** [62].

$$E = \frac{S\sqrt{\pi}}{2\sqrt{A(h)}} \quad (S2)$$

where $E$ is the Young's modulus (121.5 GPa) and $A(h)$ is the area function **equation (S1)**, both of which are assumed to be constants during the test. Since the displacement amplitude is around 1 nm within a period of 1/200 s, the dynamic displacement rate is above 200 nm/s, which is much higher than the thermal drift rate of 0.02 nm/s. Thus, the thermal drift can be ignored and the measured contact stiffness and calculated indentation depth by **equations (S1)** and **(S2)** are plotted as a function of relaxation time in **Supplementary Fig. 5c** and **d** respectively. It is noted that dynamic tests cannot measure the highest stress and/or strain rate region at the early stage of relaxation, for example 0 – 20 s, therefore the static (0 – 200 s, **Supplementary Fig. 5b**) and dynamic (20 – 2000s, **Supplementary Fig. 5d**) relaxation results are plotted together in **Supplementary Fig. 5e**, which overlap well in the time range from 20 s to 200 s and form a continuous line.

With the nanoindentation measurements, shear-strain rate ($\dot{\gamma}$) beneath the indenter is derived from the displacement rate ($dh/dt$) and depth ($h$) through an equation [63]:

$$\dot{\gamma} = 0.16 \frac{1}{h} \frac{dh}{dt} \quad (S3)$$

The hardness of the $Zr_{55}Cu_{30}Al_{10}Ni_5$ metallic glass was measured to be 6.7 GPa at room temperature which is about 3.5 times of the compression strength of 1.9 GPa [61]. Since the pure shear stress is approximately $\sqrt{3}$ times lower than the uniaxial yield stress under the monotonic loading conditions according to von Mises yield criterion, the hardness/shear strength ratio is approximately $3.5\sqrt{3}$. Thus, the nominal shear stress ($\tau$) can be estimated from indentation depth ($h$) and load ($P$) by:

$$\tau = \frac{\sigma}{\sqrt{3}} = \frac{P}{3.5\sqrt{3}A(h)}, \quad (S4)$$

where σ is the normal stress, $A(h)$ is the area function of the indenter which has been carefully calibrated before and after the relaxation tests. With the data of depth versus time in **Supplementary Fig. 5e** at constant force of 10 mN, the nominal shear stress and shear strain-rate as a function of time can be derived and are plotted in **Supplementary Fig. 5f**. The nominal shear stress beneath the indenter decreases from 1402.6 MPa ± 5 MPa to 1057.1 MPa ± 35 MPa during the relaxation from 0 s to 2000 s. Accordingly, the strain rate in relaxation testing covers more than four orders of magnitude from 0.56 $s^{-1}$ to $5.3\times10^{-6}$ $s^{-1}$ (**Supplementary Fig. 5f**). The estimated shear stress is plotted as the function of shear strain rate in **Fig. 1b**, in which the shear stress has a linear log-log relation with shear strain rate in the high stress region above 1350 MPa. The slope of the linear portion, or the strain-rate sensitivity ($m = \partial \ln \tau / \partial \ln \dot{\gamma}$), is calculated to be 0.004 ± 0.001, consistent with that of rate-jump indentation measurements [18]. The strain-rate sensitivity gradually increases to ~0.5 when the strain rate decreases to ~$3\times10^{-5}$ $s^{-1}$ and stresses is below 1100 MPa. For the broadband nanoindentation experiments, it approaches the rate limit when the relaxation extends longer than 2000 s with a resulting strain rate of $5.3\times10^{-6}$ $s^{-1}$. Beyond the time limit, the relaxation measurements will be obviously interfered by lateral thermal drift. To overcome the technical limitation, we employed a nanoindenation based cantilever bending method [64] to detect the flow behaviour of metallic glasses in a low strain rate range from $10^{-8}$ to $10^{-6}$ $s^{-1}$.

**Cantilever relaxation experiments**

Typical cantilevers with sizes are shown in **Supplementary Fig. 6a** and **b**. A conical indenter with 1.5 μm tip radius was loaded on the cantilever with a constant force of 1.0, 1.5, 2, 2.5, 3, 3.5, 4.5, 6, 8, 10 mN using a Hysitron Triboindenter. The distance between the clamp and indenter tip contact point was 400 μm for 30×70×700 μm³ cantilevers and 50μm for 5×25×200 μm³ cantilevers. The indenter displacement was measured during the constant force relaxation for 1000 seconds, and anelastic recovery was also monitored for 1000 seconds with 1 μN force. Thermal drift was measured with 1 μN contact force for 200 seconds before each test and was subtracted. Indenter displacements, or cantilever deflection displacements ($h'$), as a function of time during relaxation tests are plotted in **Supplementary Fig. 6c** and **d** for 5 μm and 30 μm thick cantilevers, respectively. The corresponding shear stresses ($\tau$) and instantaneous shear strains ($\gamma_i$) at the pivot point of cantilevers are calculated by:

$$\tau = \frac{6PL}{\sqrt{3}wb^2} \quad \text{and} \quad \gamma_i = \frac{3\sqrt{3}bh'}{2L^2}, \quad (S5)$$

where $P$ is the load, $L$ is the distance between the clamp and indenter tip contact point, $b$ is the sample thickness, $w$ is the width of the cantilever and $h'$ is the bending displacement [64]. Calculated shear strains at the pivot point under constant loading forces of 1 to 10 mN in the cantilevers are plotted as the function of time in **Supplementary Fig. 6c** and **d**. Relaxation process is readily observed in the constant force segment and anelastic recovery takes place in the unloading segment. The permanent strain is obtained by subtracting the anelastic strain ($\gamma_a$) from the total relaxation strain ($\gamma_t$) to estimate shear strain rates, $\dot{\gamma} = (\gamma_t - \gamma_a)/\Delta t$, at various stresses for a duration of $\Delta t = 1000$ s and are plotted with shear stress in **Fig. 1c**. The measured strain rates cover the range from $10^{-8}$ to $10^{-6}$ s$^{-1}$. The strain rate sensitivity ($m$) keeps increasing with the decrease of strain rate and reaches a constant value of ~1.0 when the strain rate is lower than $10^{-7}$ s$^{-1}$ in **Fig. 1c**.

**Acknowledgments:**

**Funding:** Sponsored by National Natural Science Foundation of China (Grant No. 51821001). M.C. is supported by the National Science Foundation (NSF DMR-1804320) with Dr. Judith Yang as the program director.

**Author contributions:** S.X.S designed and performed the broadband stress relaxation experiments and data process, F.Z. prepared the metallic glass samples and conducted microstructure characterization, S.X.S and M.W.C analysed the data, developed the models and wrote the paper. M.W.C conceived and supervised the project.

**Competing financial interests:** The authors declare no competing financial interests.

**Additional Information: Supplementary Information** is available for this paper.

**Data availability**

The data that support the plots and other analysis in this work are available upon request.